\documentclass[amsmath,amssymb,showpacs,preprintnumbers,twocolumn]{revtex4}
\usepackage[english,dutch]{babel}
\usepackage{graphicx}
\usepackage{dcolumn}
\usepackage{bm}
\usepackage{subfigure}
\newlength{\dfltgraph}

\newcommand \be{\begin{equation}}
\newcommand \ee{\end{equation}}
\newcommand \ba{\begin{eqnarray}}
\newcommand \ea{\end{eqnarray}}

\begin{document}
\selectlanguage{english}

\title{Spatial coupling of particle and fluid models for streamers: where nonlocality matters}
\author{Chao Li$^1$}
\author{Ute Ebert$^{1,2}$}
\author{W.J.M. Brok$^{1,2}$}
\author{Willem Hundsdorfer$^1$}
\affiliation{$^1$ CWI, P.O.~Box~94079, 1090~GB Amsterdam, The
Netherlands,} \affiliation{$^2$ Department of Applied Physics, Eindhoven University of Technology, 
The Netherlands.}
\date{\today}

\begin{abstract}
Particle models for streamer ionization fronts contain correct electron energy distributions, runaway effects
and single electron statistics. Conventional fluid models are computationally much more efficient for
large particle numbers, but create too low ionization densities in high fields. To combine their respective
advantages, we here show how to couple both models in space. We confirm that the discrepancies between
particle and fluid fronts arise from the steep electron density gradients in the leading edge of the fronts. We
find the optimal position for the interface between models that minimizes computational effort and reproduces
the results of a pure particle model.
\end{abstract}


\maketitle

Streamers generically occur in the initial electric breakdown of long gaps.
They are growing filaments of weakly ionized nonstationary plasma; they are produced by a sharp ionization front
that propagates into non-ionized matter within a self-enhanced electric field. Streamers are used in
industrial applications such as lighting~\cite{Bho2005}, gas and water
purification~\cite{Win2006, Gra2007} or combustion control~\cite{Sta2006}, and they occur in natural processes 
as well such as lightning~\cite{Rak2003,Wil2006} or transient luminous events in the upper
atmosphere~\cite{Sen1995,Pas2007}. Important recent questions concern $(i)$ propagation and branching of streamers
\cite{Ebe2006} and the role of avalanches created by single electrons, $(ii)$ the electron energy
distribution in the streamer head and the subsequent gas chemistry that is used in the above applications, as
well as $(iii)$ runaway electrons and X-ray generation, possibly in the streamer zone of lightning 
leaders~\cite{Dwy2003:3, Pas2007}. The present paper deals with the efficient simulation of these problems.

Monte Carlo particle simulations~\cite{Brok2007:2,Cha2007} model these effects as they
contain the full microscopic physics; 
the deterministic electron motion between collisions is calculated and collisions of electrons with neutrals
are treated through a Monte Carlo process with appropriate statistical weights. The particle model includes
the complete electron velocity and energy distribution as well as the discrete nature of particles. However,
a drawback of such models is that the required computation resources grow with the number of particles and eventually
exceed the CPU space of any computer. This difficulty is counteracted by using superparticles carrying
the charge and the mass of many physical particles, but superparticles in turn create unphysical fluctuations and
stochastic heating~\cite{IEEE08}.

Streamers are therefore mostly modeled as fluids (see e.g.~\cite{Mon2006:3,Seg2006,Mos2006,Liu2006,Luq2007})
since a fluid model is computationally much more efficient. In the case of a negative discharge in a
pure non-attaching gas like nitrogen, it consists of continuity equations for the densities of electrons
$n_e$ and positive ions $n_p$ coupled to the Poisson equation for the electric field ${\bf E}$.
The electron mobility $\mu$ and diffusion matrix ${\bf D}$ and
the impact ionization rate $\alpha$ are calculated from microscopic scattering and transport models like the
Boltzmann equation~\cite{Hag2005} or directly from Monte Carlo simulations as, e.g., in~\cite{Li2007:1}. In streamer
calculations, it is generally assumed that these transport and reaction coefficients are functions of the
local electric field.

\begin{figure}
\begin{center}
\includegraphics[width=0.4\textwidth,angle=270]{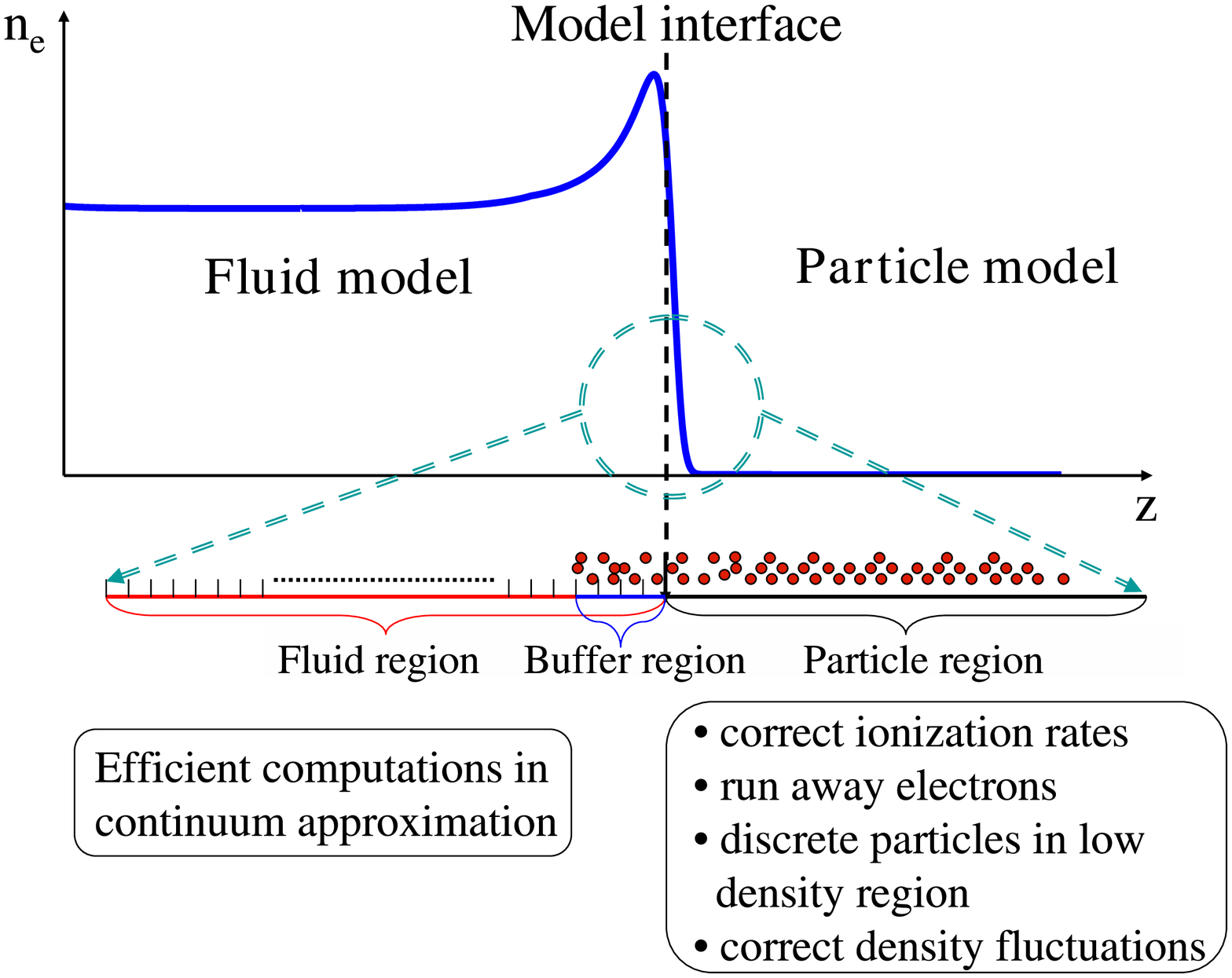}
\end{center}
\caption{\label{fig:hybrid_concept}(Color online)
The streamer ionization front, that here is indicated by the electron density $n_e(z)$,
and its presentation by particle or fluid model in different spatial regions. 
}
\end{figure}

We recently have compared the properties of streamer ionization fronts of particle models and conventional
fluid models~\cite{Li2007:1} for negative planar fronts in nitrogen; the transport coefficients
for the fluid model were generated from swarm experiments in the particle model. We found that the models
agree reasonably for fields up to 50 kV/cm at standard temperature and pressure, but that differences increase with
increasing electric field. For example, in a field of 200 kV/cm, the ionization level behind the front is
60\% higher in the particle model than in the fluid model. We have related this to the fact that the electron
energies and, consecutively, the ionization rates in the leading edge of the front are considerably higher in
the particle than in the fluid model; they are actually at the edge of runaway. We found that this effect 
is due to the strong density gradients in the front, and not due to field gradients. So for high fields 
and consecutively strong density gradients at the streamer tip, there is a clear need for particle simulations, 
and particles, rather than superparticles, should be used to get physically realistic density fluctuations 
when modeling, e.g., the branching process of a streamer.


The basic idea of the present paper is demonstrated in Fig.~\ref{fig:hybrid_concept}, namely to follow 
the single electron dynamics in the high field region of the streamer where the electron density gradient 
is steep, and to present the interior region with large numbers of slower electrons through a fluid model
with appropriate transport coefficients. As in~\cite{Li2007:1}, we study negative streamers in nitrogen,
and we simplify the notation by refering to standard temperature and pressure though the model trivially
scales with gas density. The particle and the fluid model by themselves are taken as described in
detail in~\cite{Li2007:1}. But how should particle and fluid model be coupled in space?
And where should the interface between the models be located to get fast, but reliable results? 
The answers to these questions will be given below. They required us to identify correctly 
the spatial region where particle and fluid model deviate, and this allowed us to then compute 
the full electron physics efficiently in the relevant region. 


When coupling the models, the model interface should move with the ionization front; this keeps the total
number of electrons limited and superparticles need not be introduced. The position
of the interface can be chosen either according to the electron densities or to the electric field.
As the electron densities fluctuate stronger than the electric field, we relate the position of the model 
interface to the electric field.
More precisely, the interface is placed where the local field $E$ is a given fraction $x$ of the maximal
field $E^+$: $E_{\rm interface}=x\;E^+$. By varying $x$, the region modeled by particles can be varied.

\begin{figure}
\begin{center}
\includegraphics[width=0.35\textwidth,angle=270]{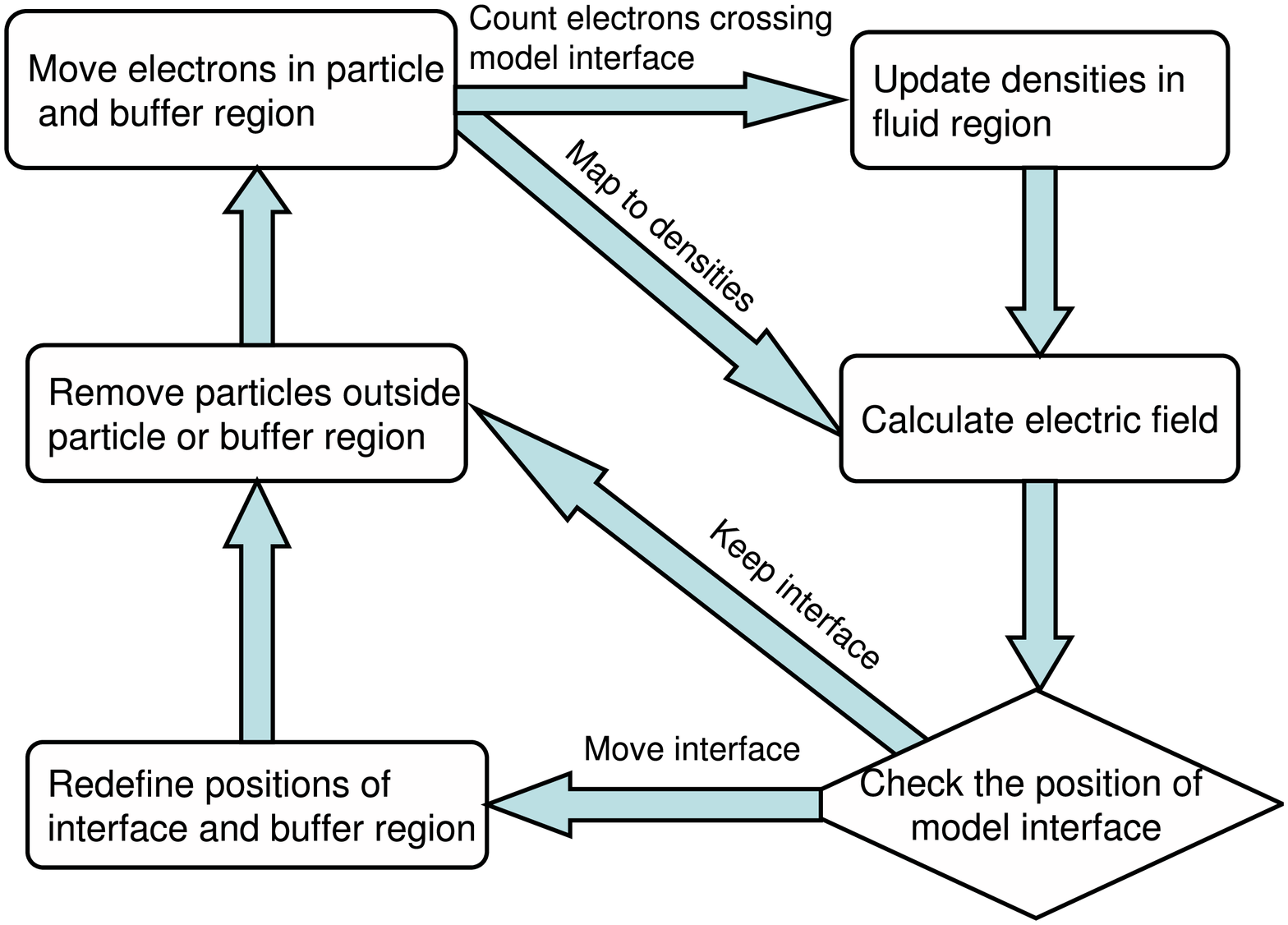}
\end{center}
\caption{\label{fig:hybrid_flow_chart}
Flow chart for one time step from $t_n$ to $t_{n+1}$ in the complete hybrid calculation.
}
\end{figure}

To properly handle the interaction of two models, we introduced a so called ``buffer region'' 
where a particle model coexists with a fluid model. 
The separation of the full computational region into fluid, particle and buffer region is indicated in Fig.~1.
Buffer regions have been introduced in~\cite{Gar1999, Ale2002, Ale2005} 
for rarefied gases coupling a direct simulation Monte Carlo (DSMC) scheme to the Navier-Stokes equations, 
and in other applications~\cite{Del2003,Ozg2002}. Physical observables are evaluated from the fluid model 
in its whole
definition region up to the model interface. Beyond that point, the particle model is used.
The particle model extends back beyond the model interface into the buffer region where particle and fluid model 
coexist; it supplies particle fluxes to the fluid model on the model interface. However, correct particle 
fluxes require correct particle statistics within the buffer region whose length should be as small as possible to reduce computation costs, but larger than the electron energy relaxation length~\cite{Li2007:1}. 
In many cases, new particles need to be introduced into the buffer region, that have to be drawn from appropriate distributions in configuration space. This would pose a particular problem since a Boltzmann or even a Druyvesteyn distribution can be inaccurate. But for negative streamers, 
where electrons on average move somewhat slower than the ionization front, the electron loss at the end of
a sufficiently long buffer region does not affect the calculation of particle fluxes at the model interface.
Therefore the particle loss at the end of the buffer region can be ignored and new electrons do not need
be created artificially. 

\begin{figure}
\begin{center}
\includegraphics[width=0.48\textwidth]{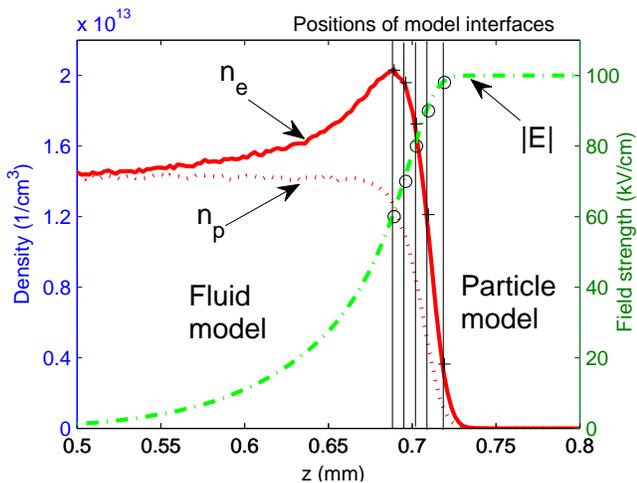}
\end{center}
\caption{\label{fig:hybridconcept}
Electron density $n_e$ and ion density $n_p$ (solid and dotted line) and electric field strength $|E|$ 
(dash-dotted line) in a streamer ionization front in nitrogen in a field of $E^+=-100$~kV/cm at
standard temperature and pressure within a pure particle model. Our units are related to other commonly used units like $1$ kV cm$^{-1}$ bar$^{-1}$ = $1.316$ V cm$^{-1}$ torr$^{-1}$ = $0.424$ Townsend at T=$300$ K.
Below we will model the leading edge of the front by a particle model and the streamer interior 
by a fluid model where the model interfaces are located at $E_{\rm interface}=x\;E^+$ with $x=0.6$, 
$0.7$, $0.8$, $0.9$, and $0.98$. These interface positions for Figs.~2 and 3 are marked by vertical lines, with ``o'' for the fields and with ``+'' for the densities.}
\end{figure}

In more detail, the calculation is performed as follows.
One hybrid computation step from $t_n$ to $t_{n+1}$ is described in the flow chart in 
Fig.~\ref{fig:hybrid_flow_chart}. The electric field $E$, the electron and ion densities $n_e$ and $n_p$ 
in the fluid region and the kinetic information of particles in the particle and buffer region are given at 
time step $t_n$. First, the positions and velocities of all old and the newly generated particles 
are updated to time step $t_{n+1}$
in the particle and in the buffer region. Their collisions during this time step are treated stochastically 
and their new velocities and positions are calculated by solving the equation of motion.
The number of electrons crossing the model interface during this time step is recorded. 
This particle flux across the interface provides the required boundary condition for calculating the evolution 
of the densities in the fluid region up to the same time $t_{n+1}$. The particles are at arbitrary positions, 
but densities and field are calculated here on the same numerical grid. Therefore the particles in 
the particle region are averaged to densities on the numerical grid, and then the electric field 
at time $t_{n+1}$ is calculated from the Poisson equation everywhere. (The charge density in the buffer 
region is taken from the fluid model, and the particles in the buffer region only serve to generate correct
fluxes for the fluid region.)
Finally, the position of the model interface is updated to its new position at time $t_{n+1}$;
it can stay where it was or move one grid size forward. The buffer region moves with it. 
All particles that now are neither in the particle nor in the buffer region, are removed from the particle list.

In the particle model, a standard PIC/MCC (Particle in Cell/Monte Carlo
collision) method is implemented. At each time step of length $\Delta
t=0.3$ ps, particles in the particle regions are mapped to
densities on a uniform grid with mesh size $\Delta \ell= 2.3$ $\mu$m.
Meanwhile, the fluid equations are solved in the fluid region of the
same grid; discretization and grid dependence of the fluid model are discussed in detail 
in~\cite{Mon2006:3}. The charge density  $n_p-n_e$ then can be obtained
everywhere and the electric field is calculated on this grid. The size
of the time step and the grid size are chosen such that the ionization
front need several time steps to move over one $\Delta \ell$, e.g., 10
$\Delta t$ at 100 kV/cm and 3 $\Delta t$ at 200 kV/cm.

The length of the buffer region is another crucial factor in the hybrid computation. A buffer region with length of 32 $\Delta \ell$ has been used in the present simulations, which ensure a reliable flux around the model interface and stable results of hybrid simulations. The length of the buffer region is much larger than the energy relaxation length found in~\cite{Li2007:1}. The long buffer region does not bring a heavy burden to the simulation of the planar front system but will considerably reduce the computational efficiency in a more complex geometry. Therefore, the minimal length of the buffer region as well as other features of fluid and particle models shall be investigated in more detail in a furture paper.

\begin{figure}
\begin{center}
\includegraphics[width=0.48\textwidth]{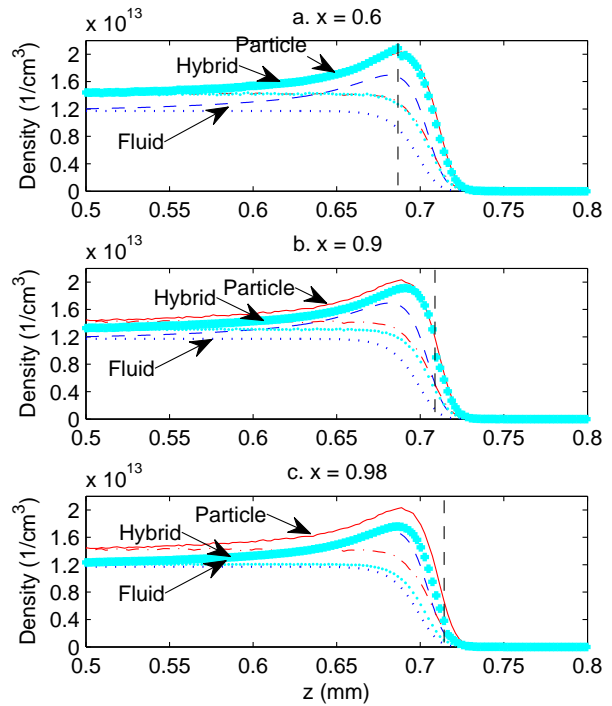}
\end{center}
\caption{\label{fig:dencom} Electron and ion
densities in the coupled model (thick lines) in a field of $E^+=-100$ kV/cm with model interfaces at $E_{\rm
interface}=x\;E^+$ with $(a)$ $x=0.6$, $(b)$ $x=0.9$, and $(c)$ $x=0.98$; these interface positions are indicated
by vertical dashed lines. The densities in the particle model
(electrons: solid, ions: dot-dashed) and in the fluid model (electrons: dashed, ions: dotted) are shown as well;
they are discussed in~\cite{Li2007:1}.}
\end{figure}

We first show the simulation results of this coupled model for a front propagating into a field of $E^+=-100$
kV/cm, and with the model interface located at $x=0.6$, $0.9$, and $0.98$; the positions of these interfaces
are indicated in Fig.~\ref{fig:hybridconcept}. The field ahead of the front is fixed, and the system is always 
taken long enough that effects at the outer boundaries are not felt.
The coupled model generates different electron and ion
densities behind the ionization front as shown in Fig.~\ref{fig:dencom}; for $x=0.6$, the density is as 
in the particle model; for $x=0.98$, it is as in the fluid model; and for $x=0.9$, it takes some intermediate 
value. We conclude that the solution of the pure particle model can be replaced by the coupled model, 
if a sufficiently large region of the ionization front with its steep gradients is covered by the particle model, 
and that the coupling to the fluid
model behind that region does not cause numerical artifacts. This confirms the discussion in~\cite{Li2007:1}; 
it is indeed the high
electron density gradient that causes an electron energy overshoot and a higher ionization rate in the
leading edge of the particle front. The coupled model also confirms that the field gradients do not play a
role in causing the density discrepancy between the fluid and the particle model as the field keeps varying across
the model interface in the coupled model.

\begin{figure}
\begin{center}
\includegraphics[width=0.5\textwidth]{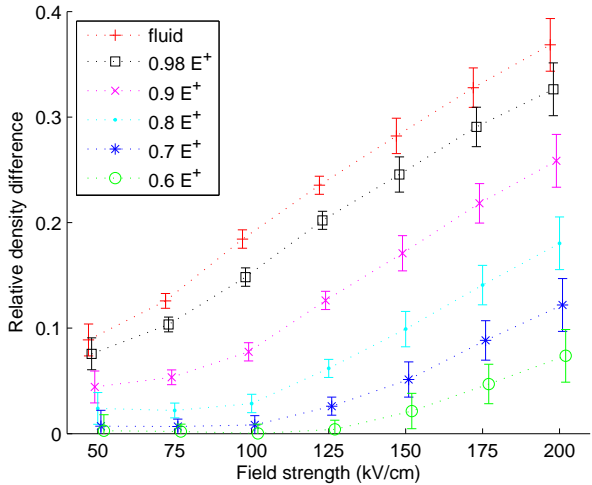}
\end{center}
\caption{\label{fig:reldiff} Relative density difference behind the front $(n_{e,part}^- -
n_{e,coup}^-)/{n_{e,part}^-}$ between the particle model and the coupled model as a function of the applied electric
field $E^+$ and of the position of the model interface $E_{\rm interface}=x\cdot E^+$ with $x=0.6$
($\square$), $0.7$ (x), $0.8$ ($\cdot$), $0.9$ (*), $0.98$ (o) and 1.0 (+). The last case corresponds to the
fluid model. (Note that a density difference of 60\% relative to the fluid model corresponds to a density
difference of 37.5\% relative to the particle model.)}
\end{figure}

Having analyzed the ionization front propagating into a field of $E^+=-100$ kV/cm, we now summarize the
results for fields ranging from $-50$ to $-200$ kV/cm.
Fig.~\ref{fig:reldiff} shows the discrepancy between particle and coupled model on the most sensitive
observable~\cite{Li2007:1}, namely on the relative difference $\left(n_{e,part}^- -
n_{e,coup}^-\right)/{n_{e,part}^-}$ of the saturated electron density $n_e^-$ behind the ionization front.
This quantity is shown as a function of the electric field $E^+$ and of the position of the model interface
parameterized again by $x$. The figure shows that for higher $E^+$, the parameter $x$ needs to be smaller. This
shift of required interface position relative to $E^+$ corresponds to a shift of the maximal electron density
relative to $E$: both for $E^+=-50$ kV/cm and for $E^+=-200$ kV/cm, the particle and the coupled model agree well, if
the model interface lies at the maximum of the electron density and therefore covers the complete steep
gradient region; this is the case at $E = 0.8\;E^+$ for $E^+=-50$ kV/cm and at $E=0.35\;E^+$ for $E^+=-200$
kV/cm.


Coupling particle and fluid models in space with varying interface positions confirms our
prediction~\cite{Li2007:1} that the density discrepancies between particle and fluid model are due to the
strong density gradients in the leading edge of the front. This investigation also lays the basis for
constructing a fully 3D coupled particle-fluid model where the fields ahead the ionization front are changing
in space and time.

{\bf Acknowledgements:} We acknowledge support by the Dutch national program BSIK, in the ICT project BRICKS, theme MSV1.



\end{document}